# Magnetic levitation force between a superconducting bulk magnet and a permanent magnet

J J Wang, C Y He, L F Meng, C Li, R S Han, Z X Gao

Department of Physics, Key Laboratory for Artificial Microstructure and Mesoscopic Physics, Peking University, Beijing 100871, P. R. China

**Abstract**

The current density $J(r,z)$ in a disk-shaped superconducting bulk magnet and the magnetic levitation force $F_z^{SBM}$ exerted on the superconducting bulk magnet by a cylindrical permanent magnet are calculated from first principles. The effect of the superconducting parameters of the superconducting bulk is taken into account by assuming the voltage-current law $E = E_c (J/J_c)^n$ and the material law $B = \mu_0 H$. The magnetic levitation force $F_z^{SBM}$ is dominated by the remnant current density $J_2'(r,z)$, which is induced by switching off the applied magnetizing field. High critical current density and flux creep exponent may increase the magnetic levitation force $F_z^{SBM}$. Large volume and high aspect ratio of the superconducting bulk can enhance the magnetic levitation force $F_z^{SBM}$ further.

1. **Introduction**

Superconducting magnets play an important part in the magnetically levitated train system. The magnetic field generated by a disk-shaped superconducting bulk magnet (SBM) can increase with its $J_c$ and volume. A large light rare earth (LRE) - $Ba_2Cu_3O_{7-x}$ (e.g., Nd, Sm, Eu, or Gd) SBM is believed to be able to trap magnetic field more than 5 T at 77 K [1-3]. Therefore, the possibility of high critical temperature SBM for the magnetically levitated system has been investigated [4-5]. The calculation of the hysteretic force between a superconductor and a permanent magnet (PM) has been reported [6]. In such calculations, however, there is no trapped magnetic field in the superconductor. We consider that the superconductor has been magnetized before a PM is placed under the superconductor, so the superconductor acts as an SBM with trapped magnetic field, and then calculate the magnetic levitation forces between the PM and the SBM.

In practice the superconducting disk can be cooled below $T_c$ in an axial applied uniform magnetic induction $B_a$, then the applied magnetic field is switched off, and the magnetic flux escapes from the disk. This is called the escape process. The calculation of the current density in the superconducting disk will be divided into three steps with different applied magnetic fields. First, we consider the surface screening current density $J_1'$ induced just after switching off the axial applied uniform magnetic field such that $\dot{B}_a \neq 0$. Second, the remnant current density $J_2'$ generated after $\dot{B}_a = 0$ and $B = 0$, which creates the trapped magnetic field of the superconducting disk. Finally, the current density $J_3$ generated by the non-uniform magnetic field of the PM. The result obtained in the previous step is taken as the initial current density in the next step. It is difficult to calculate the above current densities $J_1'$ and $J_2'$ in a superconducting disk directly, so we suggest a method that shuns such a difficulty.

We consider another process that the magnetic flux penetrates into the same superconducting disk. After the superconducting disk is cooled in zero magnetic field below its critical temperature, we start at time $t = 0$ with $B_a = 0$ and then switch on an axial applied uniform magnetic field such that $\dot{B}_a \neq 0$. The surface screening current $J_1$ is induced. As soon as $\dot{B}_a = 0$ and $B_a = \text{const}$, the superconducting disk is in the axial applied uniform magnetic field. During the penetration of the magnetic flux into the disk, the current density $J_2$ generates. This is called the penetration process.

In the above escape and penetration processes, the disk, the uniform magnetic field and the flux motion are the same, so the gradient of magnetic field has the same value but in opposite direction, that is $J_1 = -J'_1$ and $J_2 = -J'_2$. The $J_1$ and $J_2$ can be calculated from the equation of motion for current density in a superconducting disk.

## 2. Modeling

### A. Configuration

We consider a disk-shaped SBM with radius $a$ and thickness $2b$, levitated over a co-axial cylindrical PM with radius $R_{PM}$ and thickness $d_{PM}$. The center of the SBM is taken as the origin of the cylindrical coordinate system $(r,\phi,z)$, shown in Fig. 1.

### B. Basic equation for current density

We use the method presented by E. H. Brandt [6-8] to calculate the current density in a superconducting disk. The key issue is to find an equation of motion for the current density $J(\mathbf{r},t)$ inside the disk. Using the Maxwell equation $\mathbf{J} = \nabla \times \mathbf{H}$, and considering $\mathbf{B} = \mu_0 \mathbf{H}$, $\nabla \cdot \mathbf{A}_J = 0$ and $\nabla \times \mathbf{B}_a = 0$ (assuming no current sources i.e. no contacts), we obtain the equation of motion for the current density: $\mu_0 \mathbf{J} = \nabla \times \mathbf{B} = \nabla \times \nabla \times \mathbf{A}_J = -\nabla^2 \mathbf{A}_J$, where $\nabla \times \mathbf{B} = \nabla \times (\mathbf{B}_a + \mathbf{B}_J) = \nabla \times \mathbf{B}_J$ and $\nabla \times \nabla \times \mathbf{A}_J = \nabla(\nabla \cdot \mathbf{A}_J) - \nabla^2 \mathbf{A}_J$. As usual, the displacement current, which contributes only at very high frequencies, is disregarded in this "eddy-current approximation". We describe the superconductor by the material law $\mathbf{B} = \mu_0 \mathbf{H}$ and the voltage-current law $E(J) = E_c (J/J_c)^n$. $\mathbf{B} = \mu_0 \mathbf{H}$ is a good approximation when the flux density $B$ and the critical sheet current $J_c b$ are larger than the lower critical field $B_{c1}$ everywhere inside the superconducting disk [7]. This requirement is often satisfied in the magnetic levitation measurement. The voltage-current law $E(J) = E_c (J/J_c)^n$ is actually a result of the current dependence of the activation energy $U(J) = U_0 \log(J_c/J)$ by using the Arrhenius law $E = B u = B u_0 \exp[-U(J)/k_B T]$. So we obtain the parameters $E_c = B u_0$, $n = U_0 / k_B T$ and $u = u_0 \exp(-U/k_B T)$, which is the effective vortex velocity. $\sigma (= n-1)$ is the flux creep exponent.

Because of the axial symmetry the current density $\mathbf{J}$, electric field $\mathbf{E}$, and vector potential $\mathbf{A}$ (defined by $\nabla \times \mathbf{A} = \mathbf{B}$) have only one component pointing along the azimuthal direction $\hat{\phi}$; thus $\mathbf{J} = J(r,z)\hat{\phi}$, $\mathbf{E} = E(r,z)\hat{\phi}$, and $\mathbf{A} = A(r,z)\hat{\phi}$. Since $\mathbf{B} = \mu_0 \mathbf{H}$, we have $\mu_0 J = -\nabla^2 A_J$, where $A_J = A - A_f$ is the vector potential generated by the current density in the disk, and $A_f$ is the vector potential of applied magnetic field. The solution of this Poisson equation in cylindrical geometry is

$$A(\mathbf{r},z) = -\mu_0 \int_0^a d\mathbf{r}' \int_{-b}^{b} dz' Q(\mathbf{r},\mathbf{r}') J(\mathbf{r}') + A_f \qquad (1)$$

with $\mathbf{r}=(r,z)$ and $\mathbf{r}'=(r',z')$. The integral kernel

$$Q_{cyl}(\mathbf{r},\mathbf{r}') = f(r,r',z-z') \qquad (2)$$

with

$$f(r,r',z-z') = \int_0^\pi \frac{d\varphi}{2\pi} \frac{-r'\cos\varphi}{\left[(z-z')^2 + r^2 + r'^2 - 2rr'\cos\varphi\right]^{1/2}}$$

$$= \frac{-1}{\pi k}\sqrt{\frac{r'}{r}}\left[\left(1-\frac{1}{2}k^2\right)K(k^2) - E(k^2)\right] \qquad (3)$$

where

$$k^2 = \frac{4rr'}{(r+r')^2 + (z-z')^2} \qquad (4)$$

$K$ and $E$ are the complete elliptic integrals of the first and second kind, respectively.

To obtain the desired equation of motion for $J(r,z,t)$, we express the induction law $\nabla \times \mathbf{E} = -\dot{\mathbf{B}} = -\nabla \times \dot{\mathbf{A}}$ in the form $\mathbf{E} = -\dot{\mathbf{A}}$. The gauge of $\mathbf{A}$ $(\nabla \cdot \mathbf{A} = 0)$, to which an arbitrary curl-free vector field may be added, presents no problem in this simple geometry. Knowing the material law $E = E_c(J/J_c)^n$, we obtain $\dot{A} = -E(J)$. This relation between $\dot{A}$ and $J$ allows us to eliminate either $A$ or $J$ from Eq. (1). After eliminating $A$, we obtain

$$E[J(\mathbf{r},t)] = \mu_0 \int_S d^2r' Q_{cyl}(\mathbf{r},\mathbf{r}') \dot{J}(\mathbf{r}',t) - \dot{A}_F(r',z') \qquad (5)$$

This implicit equation for the current density $J(\mathbf{r},t)$ contains the time derivative $\dot{J}$ under the integral sign. In the general case of nonlinear $E(J)$, the time integration of Eq. (5) has to be performed numerically. For this purpose, the time derivative should be moved out from the integral to obtain $\dot{J}$ as an explicit functional of $J$. This inversion may be achieved by tabulating the kernel $Q_{cyl}(\mathbf{r},\mathbf{r}')$ on a 2D grid $\mathbf{r}_i$, $\mathbf{r}_j$ and then inverting the matrix $Q_{ij}$ to obtain $Q_{ij}^{-1}$, which is the tabulated reciprocal kernel $Q_{cyl}^{-1}(\mathbf{r},\mathbf{r}')$. The equation of motion for the azimuthal current density $J(r,z,t)$ then reads

$$\dot{J}(\mathbf{r},t) = \mathbf{m}_0^{-1} \int_0^a d\mathbf{r}' \int_{-b}^b dz' Q_{cyl}^{-1}(\mathbf{r},\mathbf{r}') \left[ E(J) + \dot{A}_f(\mathbf{r}',z') \right] \tag{6}$$

where $Q^{-1}$ is the reciprocal kernel, which is defined by

$$\int_0^a d\mathbf{r}' \int_{-b}^b dz' Q^{-1}(\mathbf{r},\mathbf{r}') Q(\mathbf{r}',\mathbf{r}'') = \mathbf{d}(\mathbf{r}-\mathbf{r}'') \tag{7}$$

**C. Current density in SBM**

For an axial applied uniform magnetic induction $\mathbf{B}_a = B_a \hat{z}$, we choose the vector potential $A_f = -\frac{r}{2} B_a$. At time $t = 0$ with $B_a = 0$ and $J = 0$ we switch on the applied field such that $\dot{B}_a \neq 0$, and then the surface screening current is easily induced. Immediately after that, at time $t = +e$, the magnetic field and current density inside the disk are still zero since they need some time to diffuse into the conducting material. Therefore, at $t = +e$ the electric field $E(J)$ is also zero and thus the first term in Eq. (6) vanishes. What remains is the last term, which should be the time derivative of the surface screening current $J_1$. This surface screening current is thus [7]

$$J_1(\mathbf{r},t) = -H_a(t) \int_0^a d\mathbf{r}' \int_{-b}^b dz' Q_{cyl}^{-1}(\mathbf{r},\mathbf{r}') \frac{r}{2} \tag{8}$$

The surface screening current lasts only very short time. As soon as $\dot{B}_a = 0$ and $\mathbf{B}_a = \text{const}$, the motion for the current density in the disk must be described by

$$\dot{J}_2(\mathbf{r},t) = \mathbf{m}_0^{-1} \int_0^a d\mathbf{r}' \int_{-b}^b dz' Q_{cyl}^{-1}(\mathbf{r},\mathbf{r}') \left[ E_2(J_2) \right] \tag{9}$$

Eq. (9) is easily time integrated by starting with $J_2(r,z,t_{02}) = J_1$ and then putting $J_2(r,z,t=t+dt) = J_2(r,z,t) + \dot{J}_2(r,z,t) dt$.

After about $10^2$ s the distribution of the current density in the disk reaches its saturated value, then the saturated value decreases with time. We take $J_2' = -J_2$ and the current density at $t = 1800$ s after switching off the applied magnetic field as the initial current density $J_{30}(r,z,t_{03}) = -J_2'(r,z, t=1800\text{ s})$ in next step, and the permanent magnet is placed at $-(z_{00}+2z_0)$. $z_{00}+2z_0$ is the initial distance and $z_{00}$ is the minimum distance between the top surface of the PM and the bottom surface of the SBM.

**D. The permanent magnet**

Because of the axial symmetry of the system, only the cross section of the system is considered, with the axis direction $z$ chosen as the symmetry axis of the SBM and the PM, and $\mathbf{r}$ (radial direction) parallel to the surfaces of the SBM and the PM. The top surface center of the PM approaches and recedes from the SBM as

$$s = z_0 - z_0 \sin(\mathbf{w} \cdot t) + b + z_{00} \tag{10}$$

where frequency $\mathbf{w}$ represents the speed at which the PM approaches and recedes from the SBM. Experimentally, uncertainty will be caused when the PM touches the SBM, and therefore the limit $z_{00} = 0$ should be avoided. In this calculation, we choose $z_{00}/z_0 = 0.1$ as the minimum distance between the PM and the SBM.

For this configuration, the vector potential of the PM has only one component along the $\mathbf{f}$ direction and can be derived by integrating the vector potential of a circular current loop with radius $R_{PM}$ along the thickness $d_{PM}$ as,

$$A_f(\mathbf{r}, z) = \frac{B_{rem}}{2\mathbf{p}} \int_0^{\mathbf{p}} R_{PM} \cos\mathbf{f} \ln \frac{(z+s+d_{PM}) + \sqrt{R_{PM}^2 + \mathbf{r}^2 - 2\mathbf{r}R_{PM}\cos\mathbf{f} + (z+s+d_{PM})^2}}{z+s+\sqrt{R_{PM}^2 + \mathbf{r}^2 - 2\mathbf{r}R_{PM}\cos\mathbf{f} + (z+s)^2}} d\mathbf{f} \tag{11}$$

where $B_{rem}$ is the remnant induction of the PM. The radial induction $B_r^{PM} = -\partial A_f/\partial z$ can then be written as,

$$B_r^{PM}(\mathbf{r}, z) = \frac{B_{rem}}{\mathbf{p}} \sqrt{\frac{R_{PM}}{\mathbf{r}}} \sum_{i=0}^{1} \frac{(-1)^i}{k_i} \left[ \left(1 - \frac{1}{2}k_i^2\right) K(k_i^2) - E(k_i^2) \right] \tag{12}$$

where $K$ and $E$ are complete elliptic integrals of the first and second kind, respectively. And

$$k_i^2 = \frac{4\mathbf{r} R_{PM}}{(R_{PM} + \mathbf{r})^2 + (z+s+id_{PM})^2}, \quad i = 0, 1 \tag{13}$$

**E. The SBM in a non-uniform magnetic field**

After the PM is placed under the SBM, we deal with the SBM in a non-uniform magnetic field. The equation of motion for current density in SBM is $\mathbf{m}_0 J = -\nabla^2 A_J$. The solution of this Poisson equation in cylindrical geometry

$$A_J(\mathbf{r}, z) = -\mathbf{m}_0 \int_0^a d\mathbf{r}' \int_{-b}^{b} dz' Q(\mathbf{r}, \mathbf{r}') J_3(\mathbf{r}') \tag{14}$$

and

$$\dot{J}_3(\mathbf{r}, z) = \frac{1}{\mathbf{m}_0} \int_0^a d\mathbf{r}' \int_{-b}^{b} dz' Q^{-1}(\mathbf{r}, \mathbf{r}') \left[ E_3(\mathbf{r}, \mathbf{r}') + \dot{A}_f(\mathbf{r}', z') \right], \tag{15}$$

Eq. (15) is easily time integrated by starting with $J_3(\mathbf{r}, z, t_{03}) = J_2(\mathbf{r}, z, t = 1800 \text{ s})$ and then putting $J_3(\mathbf{r}, z, t = t + dt) = J_3(\mathbf{r}, z, t) + \dot{J}_3(\mathbf{r}, z, t) dt$.

**F. The levitation force**

As the current density $J_3(r,z,t)$ and the radial magnetic induction $B_r^{PM}$ inside the SBM have been derived, the vertical levitation force along the $z$-axis can be readily obtained as

$$F_z = 2\pi \int_0^a d\mathbf{r} \int_{-b}^{b} dz J_3(\mathbf{r},z) B_r^{PM}(\mathbf{r},z) \qquad (16)$$

**3. Results and discussions**

Now we consider the value of the applied magnetic field $H_a$ in field cooling. For short cylinders in the Bean limit, we have an explicit expression for the field of full penetration $H_p$, i.e. the value of the applied magnetic field $H_a$ at which the penetrating flux and current fronts have reached the specimen center [7]. At $H_a > H_p$, the current density in the Bean model does not change anymore, and the trapped magnetic field created by the current density will not change either. For cylinders with arbitrary aspect ratio $b/a$, the field of full penetration is

$$H_p = J_c b \ln\left[\frac{a}{b} + \left(1 + \frac{a^2}{b^2}\right)^{1/2}\right] \qquad (17)$$

For $b/a = 1$, $H_p = J_c b \ln(1+\sqrt{2})$ and $B_p = \mu_0 J_c b \ln(1+\sqrt{2})$. Typically $a = 0.025$ m and $J_c = 2\times 10^4$ A/cm$^2$, then $B_p = 5.538$ T. The typical value of $B_{rem}$ for NdFeB rare-earth permanent magnet is about 0.5 T and the typical magnetic field applied to magnetize the melt-textured-grown YBCO disk is 1~2 T at 77 K, so it is reasonable to take $B_{rem}/B_p = 0.1$ and $H_a/H_p = 0.3$. We take $E_c = \mu_0 = J_c = 1$, $w = 0.1$, $s = 20$, $b/a = 1$, $R_{PM}/a = 1$, $d_{PM}/a = 1$, $z_0/a = 0.5$, $z_{00}/z_0 = 0.1$, $B_{rem}/B_p = 0.1$ and $H_a/H_p = 0.3$ in the following calculation, unless special declaration.

**A. Surface screening current $J_1$**

Fig. 2 shows the profile of the surface screening current density $J_1(r,z)$ calculated by Eq. (8). The thickness of this current-carrying layer depends on the choice of the computational grid in the disk. The layer thickness may be reduced, and the precision of the computed surface screening current can be enhanced, by choosing a non-equidistant grid, which is denser near the disk surface. An appropriate choice of such a non-equidistant grid $\mathbf{r}_i = (r_i, z_i)$ is obtained by taking $r = r(u) = \frac{1}{2}(3u - u^3)a$, and then tabulating $u = 0, \cdots, 1$ on equidistant grids $u_k = (k - \frac{1}{2})/N_r$ $(k = 1, \cdots, N_r)$. We divide $z$ into $N_z$ parts in the similar way, yielding a 2D grid of $N = N_r N_z$. The weights at different points should be considered. In this calculation we take $N \approx 400$.

A very interesting feature shown in Fig. 2 is that the screening current density in the side surface of the disk is much higher than critical current density $J_c$, especially on the brims of the top and bottom surfaces of the disk. Inside the disk $J_1(r,z)$ is almost zero. The screening current lasts only very short time. If the cooling of the surface of the disk is ideal, the dissipation heat cannot drive the disk to normal state, so the disk is still in superconducting state.

## B. Current $J_2$ profiles in SBM

As soon as $\dot{B}_a = 0$ and $B_a = \text{const}$, the surface screening current density $J_1(r,z)$ rapidly decreases. Fig. 3 shows the profiles of the current density $J_2(r,z)$ calculated by Eq. (9). $J_2(r,z)$ at the side surface of the disk rapidly decreases to a saturated value $0.85 J_c$ at $t = 0.05$ s, however, $J_2(r,z)$ inside the disk increases. After about 144 s, $J_2(r,z)$ in the whole disk reaches the same value of $0.6 J_c$. The distribution of $J_2(r,z)$ will keep this plateau feature in future, however, the height of the plateau will decrease with time very slowly. At $t = 1800$ s, $J_2(r,z)$ decreases to $0.5 J_c$.

## C. Current $J_3$ profiles in the SBM

As soon as a PM with $B_{rem} = 0.1 B_p$ is co-axially placed under the SBM at $-(z_{00} + 2z_0)$, an axial non-uniform magnetic field is applied. The initial current density in SBM takes $J_2(r,z,t = 1800 \text{ s})$, different from that one in a superconductor without trapped magnetic field, in which case the initial current density is zero.

When the PM approaches the SBM with the initial current density $J_3(r,z,t_{03}) = J_2(r,z,t = 1800 \text{ s})$, $J_3(r,z)$ at the brims of both the top and bottom surfaces of the disk increases, but the increase at the top surface is smaller. For example, at $t = 4.3$ s $J_3(r,z)$ at the brims of the bottom surface increases to $0.7 J_c$, while $J_3(r,z)$ at the brims of the top surface increases to $0.55 J_c$.

As the PM is moving closer, in most volume of the SBM $J_3(r,z)$ still equals to $J_2'(r,z)$ and in rest volume at the brims of the bottom surface $J_3(r,z)$ becomes $J_3'(r,z)$, which is determined by the non-uniform magnetic field of PM. $J_3'(r,z)$ reaches a new plateau of $0.75J_c$ at the minimum distance between PM and SBM. When the PM is moving away from the SBM, $J_3'(r,z)$ is reversed to $-0.75J_c$ at the maximum distance between PM and SBM. Up to this moment the motion of the PM completes one cycle.

**D. Comparison of levitation forces**

The vertical levitation force $F_z^{SBM}$ (solid lines) between a PM and an SBM in Fig. 5(a) shows typical hysteretic behavior. Several interesting features, much different from $F_z^{SC}$ (broken lines) between a PM and a superconductor without trapped magnetic field, are easily observed. First, as soon as a PM is placed at $-(z_{00}+2z_0)$, $F_z^{SC}=0$, but $F_z^{SBM}>0$, indicating an initial repulsive force. Second, $F_z^{SC}$ is attractive as the PM is moving away from SC, especially for higher ratio of $B_{rem}/B_p$. As the PM is moving away from the SBM, however, $F_z^{SBM}$ always shows repulsive feature. There is no attractive force in the process. Third, $F_z^{SBM}$ is much larger than $F_z^{SC}$, typically as the PM is at its equilibrium position of $z=-(z_{00}+z_0)$. The magnetic levitation force $F_z^{SC}$ is determined by the current density induced by the non-uniform magnetic field of the PM. The magnetic levitation force $F_z^{SBM}$, however, is dominated by the current density $J_2'$, which is induced by switching off the applied magnetizing field.

When the remnant induction of the PM increases to $B_{rem}/B_p=1$, both levitation forces $F_z^{SBM}$ and $F_z^{SC}$ enhance. Fig. 5(b) shows the first two cycles of $F_z^{SBM}$ and $F_z^{SC}$. The attractive part of $F_z^{SC}$ extends to longer range, however, $F_z^{SBM}$ still keeps repulsive.

**E. Effect of the aspect ratio $b/a$ on $F_z^{SBM}$**

Fig. 6 shows the vertical magnetic levitation force $F_z^{SBM}$ versus the distance $s$ for different aspect ratios $b/a$ of the superconducting disk for one cycle. The whole curve moves to higher value of $F_z^{SBM}$ with the increase of the aspect ratio $b/a$. The inset of Fig. 6 shows the maximum levitation force as a function of the aspect ratio $b/a$. It can be seen that for small aspect ratio $b/a$, the maximum levitation force $F_z^{SBM}$ increases with $b/a$ rapidly, and saturates at $b/a\approx 2$, different from $F_z^{SC}$ for $b/a\approx 1$. Technically, a superconducting disk with its diameter $2a$ approximately equal to its half thickness *b* may be optimum for magnetic levitation; further increase of its thickness will only increase its weight without enhancing the levitation force significantly.

**F. Effect of superconducting parameters on $F_z^{SBM}$**

Two superconducting parameters may influence $F_z^{SBM}$. One is the flux creep exponent $s$ related to the pinning potential. Usual value for $s$ in conventional superconductors is 150, for high $T_c$ superconductors $s$ is extrapolated to be 20~60 [9]. Smaller $s$ means lower pinning potential or higher temperature. The effect of $s$ for $s = 2, 6, 10, 20, 30, 50$ on the magnetic levitation force versus *s* is shown in Fig. 7. The whole curve moves to higher value of $F_z^{SBM}$ with the increase of $s$.

The other superconducting parameter, which drastically influences the magnetic levitation force, is the critical current density $J_c$ of the superconductor. The calculated results of $F_z^{SBM}$ versus *s* for different critical current densities are plotted in Fig. 8. The inset of Fig. 8 shows the dependence of the maximum levitation force as a function of the $J_c$.

In the previous calculation [7], the dependence of the maximum levitation force between a superconductor without trapped magnetic field and a PM as a function of critical current density is linear at very low critical current density, and saturates at high critical current density. A fitting to the calculated data results in

$$F_z = \frac{3.3 J_c}{9.3 + J_c}. \tag{18}$$

Much different from the previous calculation, the dependence of the maximum levitation force between a superconductor with trapped magnetic field and a PM as a function of critical current density increases with $J_c^a$. A fitting to the calculated data results in

$$F_z \propto J_c^a \quad (1 <?\ ? < 2), \tag{19}$$

which is shown as solid line in the inset of Fig. 8.

## 4. CONCLUSIONS

The remnant current density $J_2'(r,z,t)$ and the current density $J_3(r,z,t)$ in the SBM with trapped magnetic are calculated from first principles. From the derived current density inside the disk, the magnetic levitation force $F_z^{SBM}$ between the SBM and the PM has been determined. The superconductor is described by the material law $B = \mu_0 H$, and the flux creep is described by the voltage-current law $E = E_c (J/J_c)^n$. The magnetic levitation force $F_z^{SC}$ is determined by the current density induced by the non-uniform magnetic field of the PM; the magnetic levitation force $F_z^{SBM}$, however, is dominated by the remnant current density $J_2'(r,z)$, which is induced by switching off the applied magnetizing field. High critical current density and the flux creep exponent may increase the magnetic levitation force $F_z^{SBM}$. Large volume and aspect ratio of the superconducting disk can enhance the magnetic levitation force $F_z^{SBM}$ further. The magnetization of the superconducting disk may effectively enhance the levitation force between the superconducting disk and the PM.

## ACKNOWLEDGMENTS

This work was supported by the National Science Foundation of China (NSFC 10174004) and the Ministry of Science and Technology of China (Project No. NKBRSF-G1999064602).

Figure captions:

Fig. 1: Configuration of a SBM levitated over a PM. The center of the SBM is taken as the origin of the cylindrical coordinate system.

Fig. 2: Profile of the surface screening current density $J_1(r,z)$ calculated by Eq. (8).

Fig. 3: Current profiles $J_2(r,z)$ in the same disk at different moments.

Fig. 4: Current profiles $J_3(\mathbf{r},z)$ in the same disk for different time in the first cycle.

Fig. 5: Comparison of $F_z^{SBM}$ and $F_z^{SC}$ with different remnant induction of the PM in the first two cycles. (a) $B_{rem}/B_p = 0.1$, (b) $B_{rem}/B_p = 1$.

Fig. 6: The vertical magnetic levitation force $F_z^{SBM}$ versus the distance $s$ for different aspect ratios $b/a$ of the superconducting disk for one cycle. Inset shows the maximum levitation force as a function of $b/a$, the solid line is a guide for eyes only.

Fig. 7: The vertical magnetic levitation force $F_z^{SBM}$ versus the distance $s$ at chosen parameters for different creep exponents $s$. Inset shows the maximum levitation force as a function of $s$, the solid line is a guide for eyes only.

Fig. 8: The vertical magnetic levitation force $F_z^{SBM}$ as a function of the distance $s$ for different critical current densities $J_c$. Inset shows the maximum levitation force as a function of $J_c$, the solid line is a fit with $F_{z,\max}^{SBM} \propto J_c^a$.

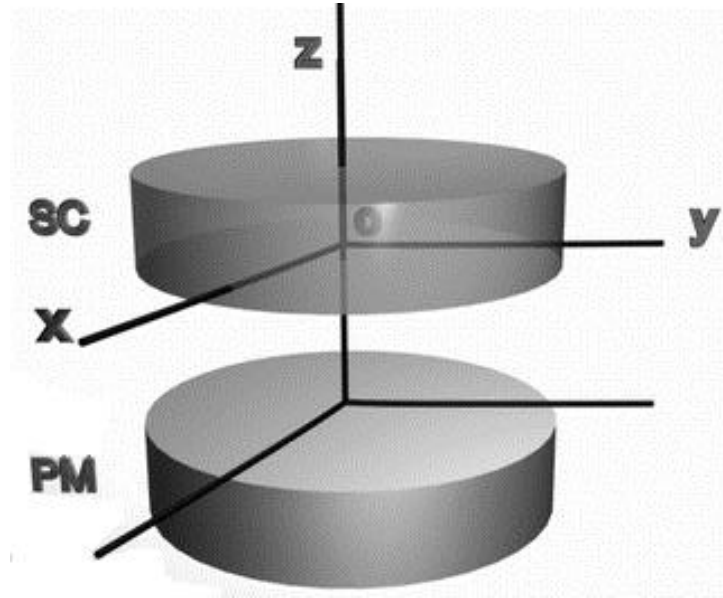

Fig. 1

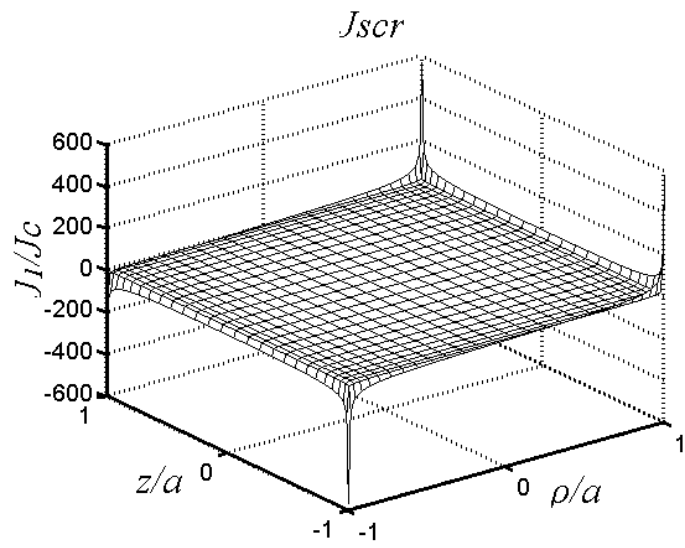

Fig. 2

$t=0.0546788$s

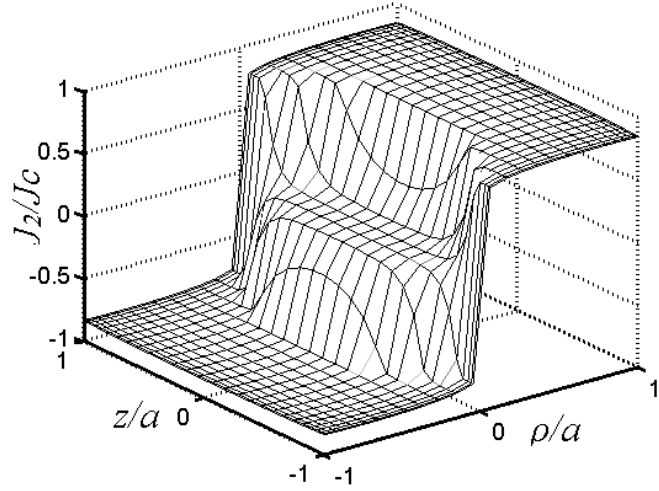

$t=144.081$s

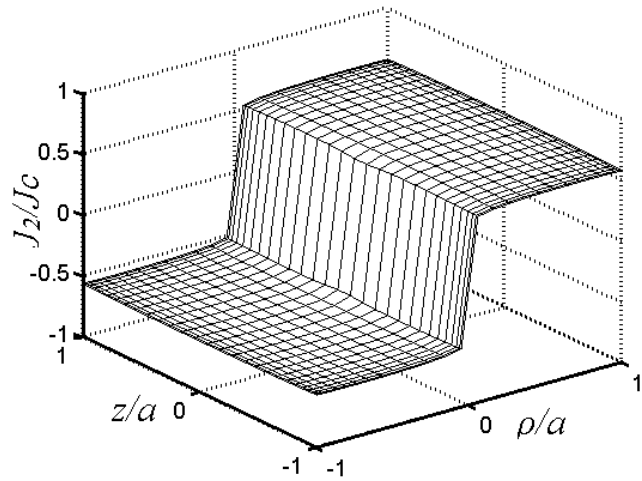

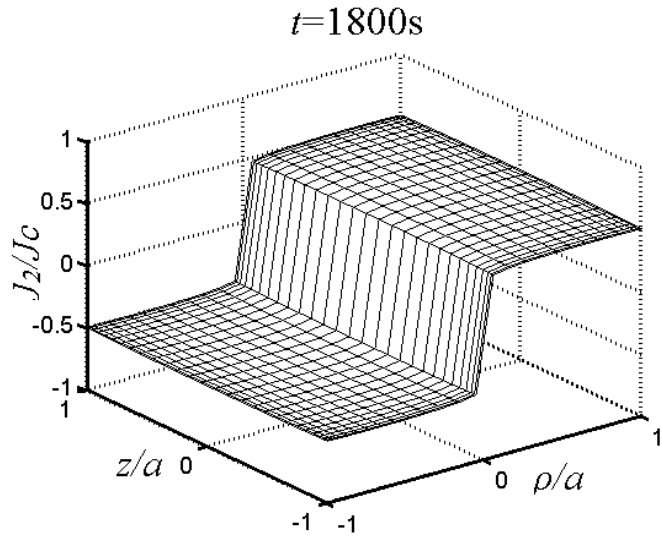

Fig. 3

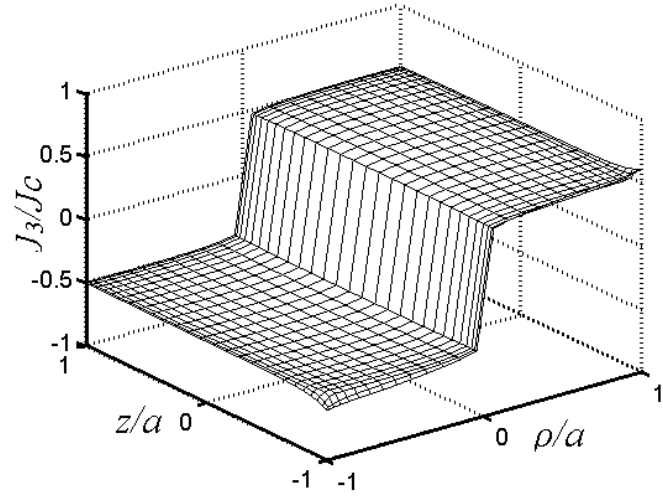

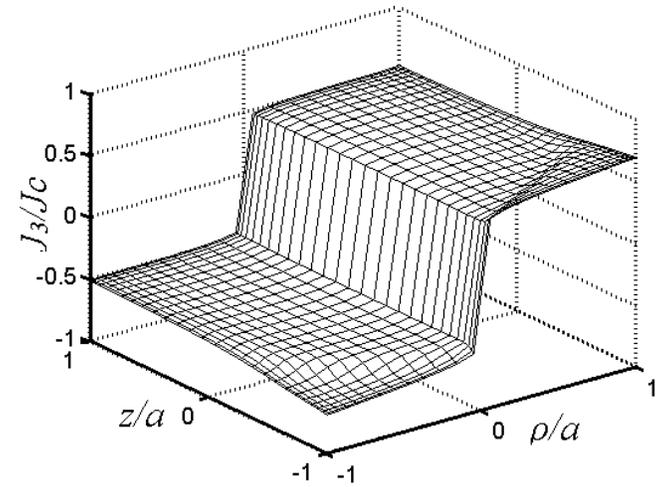

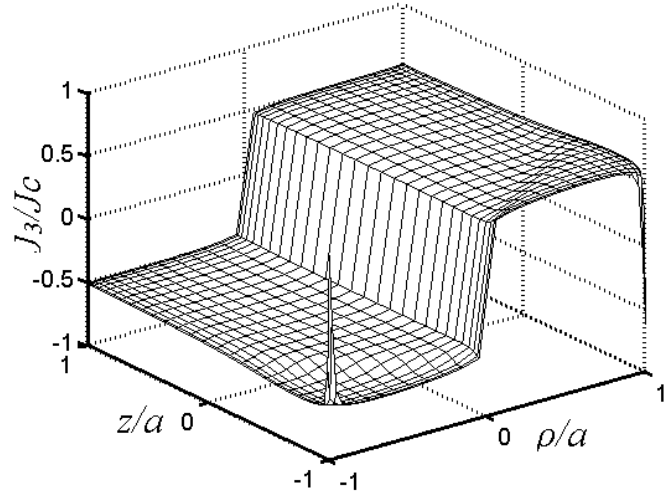

*t*=33.8331s *s/a*=0.0322677

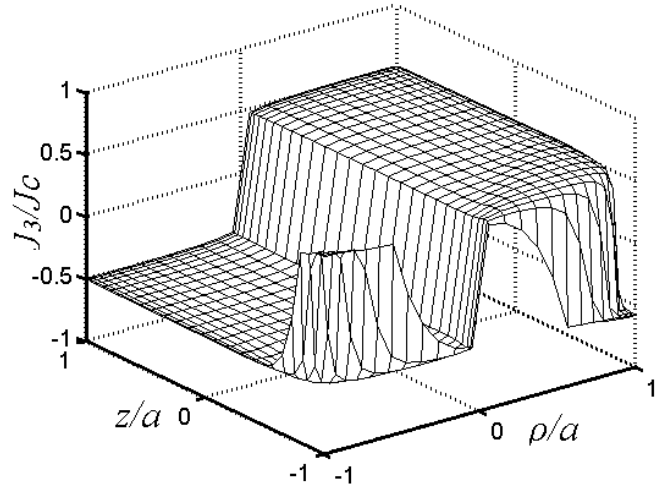

*t*=62.6071s *s/a*=0.524937

Fig.4

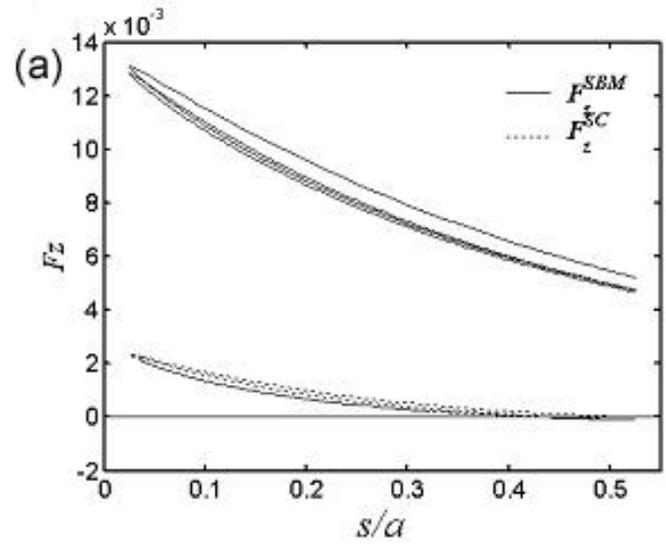

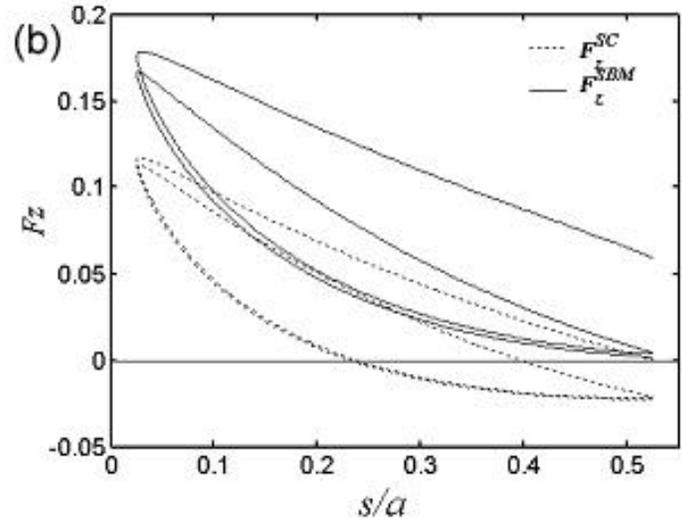

Fig. 5

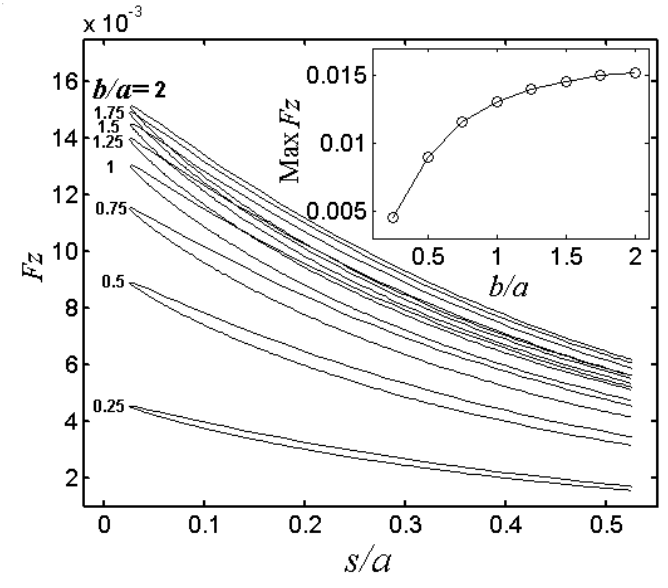

Fig. 6

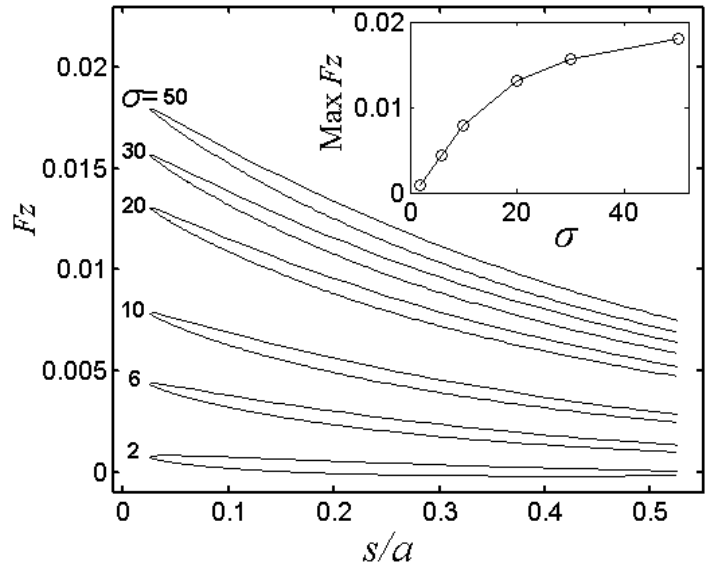

Fig. 7

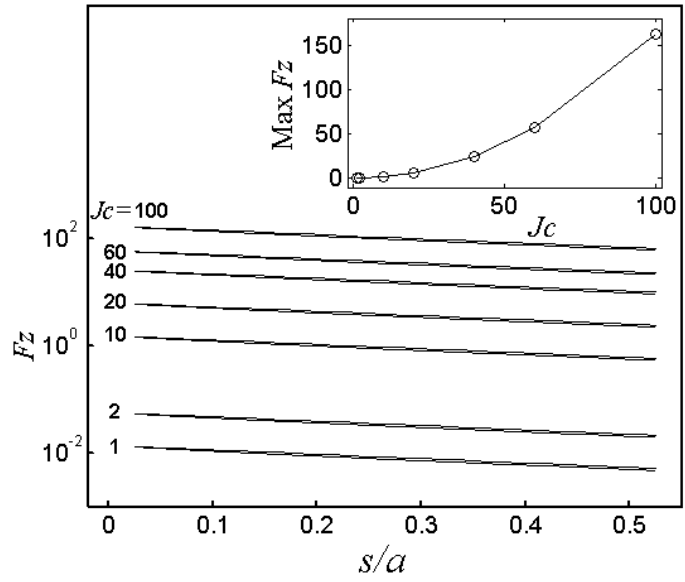

Fig. 8